\documentclass[aps,prl,reprint,twocolumn,showpacs,superscriptaddress]{revtex4-2}

\usepackage{amsmath}
\usepackage{graphicx}
\usepackage{lmodern}
\usepackage{amsmath}
\usepackage{color}
\usepackage{amssymb}
\usepackage{bm}
\usepackage{braket}
\usepackage{mathtools}
\usepackage{afterpage}
\newcommand{\spinup}{\uparrow}
\newcommand{\spindown}{\downarrow}

\newcommand{\br}{\bm{r}}
\newcommand{\bk}{\bm{k}}
\newcommand{\bq}{\bm{q}}
\newcommand{\bQ}{\bm{Q}}

\DeclareMathOperator{\sgn}{sgn}

\newcommand{\Wu}[1]{\textcolor{red}{#1}}
\newcommand{\beginsupplement}{%
   \setcounter{equation}{0}
   \pagebreak
\onecolumngrid
\setcounter{figure}{0}
\setcounter{table}{0}
\setcounter{page}{1}%
\renewcommand{\theequation}{S\arabic{equation}}%
\renewcommand{\thefigure}{S\arabic{figure}}
}
\newcommand{\be}{\begin{equation}}
\newcommand{\ee}{\end{equation}}

\usepackage[dvipsnames]{xcolor}
\usepackage[colorlinks=true]{hyperref}
\hypersetup{
    colorlinks=true,
    linkcolor=blue,
    citecolor=blue,
    urlcolor=blue
} 

\makeatletter
\def\maketitle{
\@author@finish
\title@column\titleblock@produce
\suppressfloats[t]}
\makeatother

\begin{document}
\title{Pair density wave order from electron repulsion}

\author{Yi-Ming Wu}
\affiliation{Stanford Institute for Theoretical Physics, Stanford University, Stanford, California 94305, USA}
\author{P. A. Nosov}
\affiliation{Stanford Institute for Theoretical Physics, Stanford University, Stanford, California 94305, USA}
\author{Aavishkar A. Patel}
\affiliation{Center for Computational Quantum Physics, Flatiron Institute, New York NY 10010, USA}
\author{S. Raghu}
\affiliation{Stanford Institute for Theoretical Physics, Stanford University, Stanford, California 94305, USA}

\begin{abstract}
	
	A  pair density wave (PDW) is a superconductor whose order parameter is a periodic function of space, without an accompanying spatially-uniform component.  Since PDWs are not the outcome of a weak-coupling instability of a Fermi liquid, a generic pairing mechanism for PDW order has remained elusive.  
	We describe and solve models having robust PDW phases.  To access the intermediate coupling limit,  we invoke  large $N$ limits of Fermi liquids with repulsive BCS interactions that admit saddle point solutions.  	We show that the requirements for long range PDW order are that the repulsive BCS couplings must be non-monotonic in space and that their strength must exceed a threshold value.    We obtain a phase diagram with both finite temperature transitions to PDW order, and a $T=0$ quantum critical point, where non-Fermi liquid behavior occurs.  
\end{abstract}

\maketitle

\date{\today}

{\it Introduction.  } 
A pair density wave (PDW) is a rare and exotic superconductor in which pairs of electrons condense with non-zero center of mass momentum\cite{Agterberg}.  Similar phases of matter were conceived decades ago by Fulde, Ferrel, Larkin and Ovchinnikov (FFLO), in the context of spin-polarized superconductivity\cite{FF,LO,cryst8070285,Matsuda2007,Gurevich,FeSC}.   In addition to exhibiting the usual properties of superconductors, PDWs break  translation symmetry and are therefore accompanied by charge modulation.  PDW order is believed to occur in a variety of correlated electron materials\cite{Fradkin2015,Himeda2002,Yang_2009,Raczkowski2007,Berg2007,Capello2008,PALee2014,Wang2015a,Wang2015b,Edkins2019,Wang2018}, in cold atom systems\cite{Partridge2006,Zwierlein2006,Radzihovsky2010}, and in some systems with nested Fermi surfaces\cite{Cho2012,Bednik2015,Wang2016,LiYi2018,PDWnc,Zhengzhi}.  More recently, they have been observed in the Iron based superconductor EuRbFe$_4$As$_4$\cite{Zhao2022}  as well as the Kagome metal CsV$_3$Sb$_5$\cite{Chen2021}.  Since PDWs do not stem from a weak-coupling instability of a Fermi liquid, robust  mechanisms of PDW formation have remained elusive, despite intense efforts\cite{Bhattacharyya_1995,PhysRevB.96.224503,PhysRevB.98.224501,Berg2010,Loder2010,Loder2011,PALee2014,Setty2021,Setty2022,Jiang2022}   .

It is easy to see why PDW order requires intermediate coupling.  In a clean Fermi liquid with inversion and/or time-reversal symmetry, the static pair susceptibility is a positive-definite quantity that  diverges logarithmically only at zero center of mass momentum $\bm q=0$, reflecting the celebrated BCS instability.  Away from $\bm q=0$, the logarithmic divergence is cut off, and pairing with $\bm q \ne 0$ requires a finite interaction strength.    Therefore, many proposed mechanisms for FFLO superconductivity have relied on shifting the large  pair susceptibility away from $\bm q=0$, say by the application of a zeeman field\cite{FF,LO}, or, say, by considering the effects of Rashba spin-orbit effects on odd parity superconductivity\cite{Yu2022}.  By contrast, we wish to ask whether there can be an {\it intrinsic} mechanism for PDW order, which 
requires only the existence of sizeable  interactions.  
\begin{figure}
	\includegraphics[width=7.5cm]{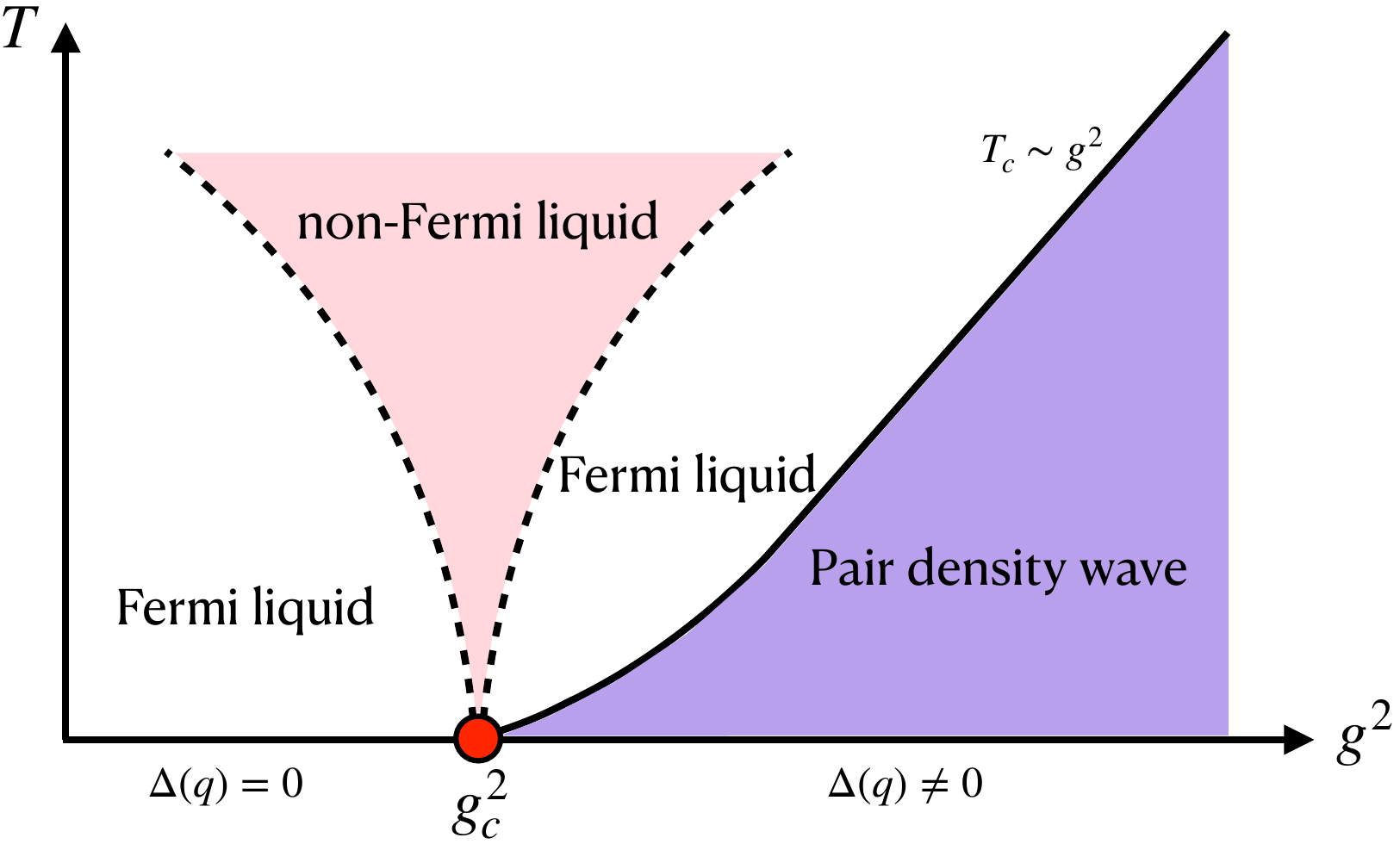}
	\caption{Phase diagram obtained from the large-$N$ model. At $T=0$, there is a QCP separating the PDW phase and the normal metallic state. The PDW transition temperature $T_c$ scales linear in $g^2$ in strong coupling limit. Above the QCP, fluctuation of PDW gives rise to NFL behavior.}\label{fig:phase}
\end{figure}

In this letter, we study various models of Fermi liquids in the presence of repulsive BCS interactions.  We solve such theories beyond the weak-coupling regime by appealing to a large $N$ limit whose saddle point corresponds to a self-consistent set of solutions for the propagators of the theory.  From these solutions, we deduce the existence of both finite temperature continuous transitions to PDW order, as well as a quantum critical point (QCP) at $T=0$ separating a 
Fermi
liquid metal from a PDW.  Our analysis leads to robust pairing mechanisms in $d>1$ of PDW order in a variety of continuum and lattice systems.  Despite such robustness, we find that PDW order emerges from physically reasonable microscopic models only under special  circumstances, which we precisely outline below.  This perhaps accounts in part for why PDW order is so rare in real materials.

{\it Model and method of solution. }
We will study the fate of a Fermi liquid subject to a finite repulsive singlet BCS interaction:
\begin{equation}
H_{\text{pair}}=\sum_{ij}V_{ij}b_i^\dagger b_j, ~b_i=c_{i\spindown} c_{i\spinup},\label{eq:ppI}
\end{equation}
In a translationally invariant system, $V_{i j} = V(\bm r_i - \bm r_j )$, and the interaction above can equivalently be expressed in momentum space as $H_{\text{pair}} = \sum_{\bm q} V(\bm q) b^{\dagger}_{\bm q} b_{\bm q}$.
We decouple the above interaction using an auxiliary field $\phi$, which corresponds to a charge $2e$ pair field.  The bare euclidean Lagrangian density then consists of the metal, the pair fields, and a Yukawa coupling between them: $\mathcal L = \mathcal L_f + \mathcal L_b + \mathcal L_g$, where
\begin{eqnarray}
\mathcal L_f &=& \sum_{\sigma = \uparrow, \downarrow} \int_y \psi_{\sigma}^{\dagger}(x) G_0^{-1}(x-y) \psi_{\sigma}(y), \nonumber \\
\mathcal L_b &=& \int_y \phi^{\dagger}(x) D_0^{-1}(x-y) \phi(y), \nonumber \\
\mathcal L_g &=& \eta g \left( \phi^{\dagger}(x) \psi_{\uparrow}(x)\psi_{\downarrow}(x)  + \phi(x) \psi_{\downarrow}^{\dagger}(x) \psi_{\uparrow}^{\dagger}(x)  \right),\label{eq:Lag}
\end{eqnarray}
$x=\left( \bm x, \tau \right)$, $\eta = 1(i)$ corresponds to attractive (repulsive) BCS couplings parametrized by a dimensionless coupling $g$ (for the repulsive case, see \cite{SM} for details), and $G_0, D_0$ are respectively the bare fermion and boson propagators in the decoupled limit $g=0$ (i.e. $D_0$ is proportional to the Fourier transform of the inverse $[V(\mathbf{q})]^{-1}$).   

The theory above can be solved for arbitrary $g$ by considering a formal extension  to large $N$ limit where the  fermion and boson fields are promoted to $N$ component vectors that transform in the fundamental representation of a global $SU(N)$ flavor symmetry group.   The coupling between the fields is promoted to an all-to-all random Yukawa coupling in the space of flavors:

\begin{equation}
	\begin{aligned}
		\mathcal L_g \rightarrow  \eta \sum_{\textcolor{black}{km\ell}} \left( \frac{g_{\textcolor{black}{km\ell}}}{N} \psi_{\textcolor{black}{k} \uparrow}(x) \psi_{\textcolor{black}{m} \downarrow}(x) \phi^{\dagger}_{\textcolor{black}{\ell}}(x)\right.\\
		\left. + \frac{g^*_{\textcolor{black}{km\ell}}}{N} \psi^{\dagger}_{\textcolor{black}{m} \uparrow}(x) \psi^{\dagger}_{\textcolor{black}{k} \downarrow}(x) \phi_{\textcolor{black}{\ell}}(x)   \right),
	\end{aligned}
\end{equation}
where the quenched random Yukawa couplings are spatially independent, and are chosen from a Gaussian unitary ensemble with variance $\overline{g_{km\ell} g^*_{k'm'\ell'}} = g^2 \delta_{kk'} \delta_{mm'} \delta_{\ell\ell'}$ 
\textcolor{black}{and with zero average}. The global $SU(N)$ symmetry is thus only preserved on average. \textcolor{black}{In terms of the original fermionic operators, this extension corresponds to the interaction of the form  
\begin{equation}
H_{\text{pair}}=\sum_{ij}V_{ij}\sum\limits_{\ell}b_{\ell i}^\dagger b_{\ell j}, \;\;b_{\ell i}=\sum\limits_{km}\frac{g_{km\ell }}{N}c_{ki\spindown} c_{mi\spinup}.\label{eq:ppUI}
\end{equation}}
Using by now standard saddle point methods\cite{Esterlis2019,Esterlis2021,Aldape2022,Patel2018,Wang2020}, the exact solution of the large N theory consists of self-consistent propagators, $G, D$ with associated self-energies $\Sigma, \Pi$:
\begin{equation}
	\begin{aligned}
		\Sigma(k)&=-g^2\sum_{q}\sgn[V(\bq)]G(-k+q)D(q),\\
		\Pi(q)&=-g^2\sgn[V(\bq)]\sum_{k}G(k)G(-k+q),\\
		 G(k)&=\left[G_0^{-1}(k)+\Sigma(k)\right]^{-1},~D(q)=\left[D_0^{-1}(q)-\Pi(q)\right]^{-1}.
	\end{aligned}\label{eq:saddle}
\end{equation}
Here, $k=(\bk,i\omega_n)$ and $q=(\bq,i\Omega_m)$, where $\omega_n (\Omega_m)$ are fermion(boson) Matsubara frequencies. 
The sign function $\sgn[V(\bq)]$ originates from the factor $\eta$ introduced in Eq.\eqref{eq:Lag}.
From the exact propagators $G, D$, we extract all the salient physics, to obtain the schematic phase diagram in Fig. \ref{fig:phase}.  For instance, to identify the finite temperature PDW transitions shown in Fig. \ref{fig:phase}, we need only consider the static bosonic propagator $D(\bm q)$. The effective Ginzburg-Landau theory for the fields $\phi$ will have a quadratic term whose coefficient is given by $D^{-1}(\bm q)$.  To study the manner in which the order parameter grows below the PDW transition, we again study the static bosonic propagators but now with the inclusion of non-linear effects stemming from a non-zero vacuum expectation value of $\phi$.  Finally, we will describe the PDW QCP and find the non-Fermi liquid behavior for the fermions.

{\it Fluctuating PDW order.  } 
\begin{figure}
	\includegraphics[width=6cm]{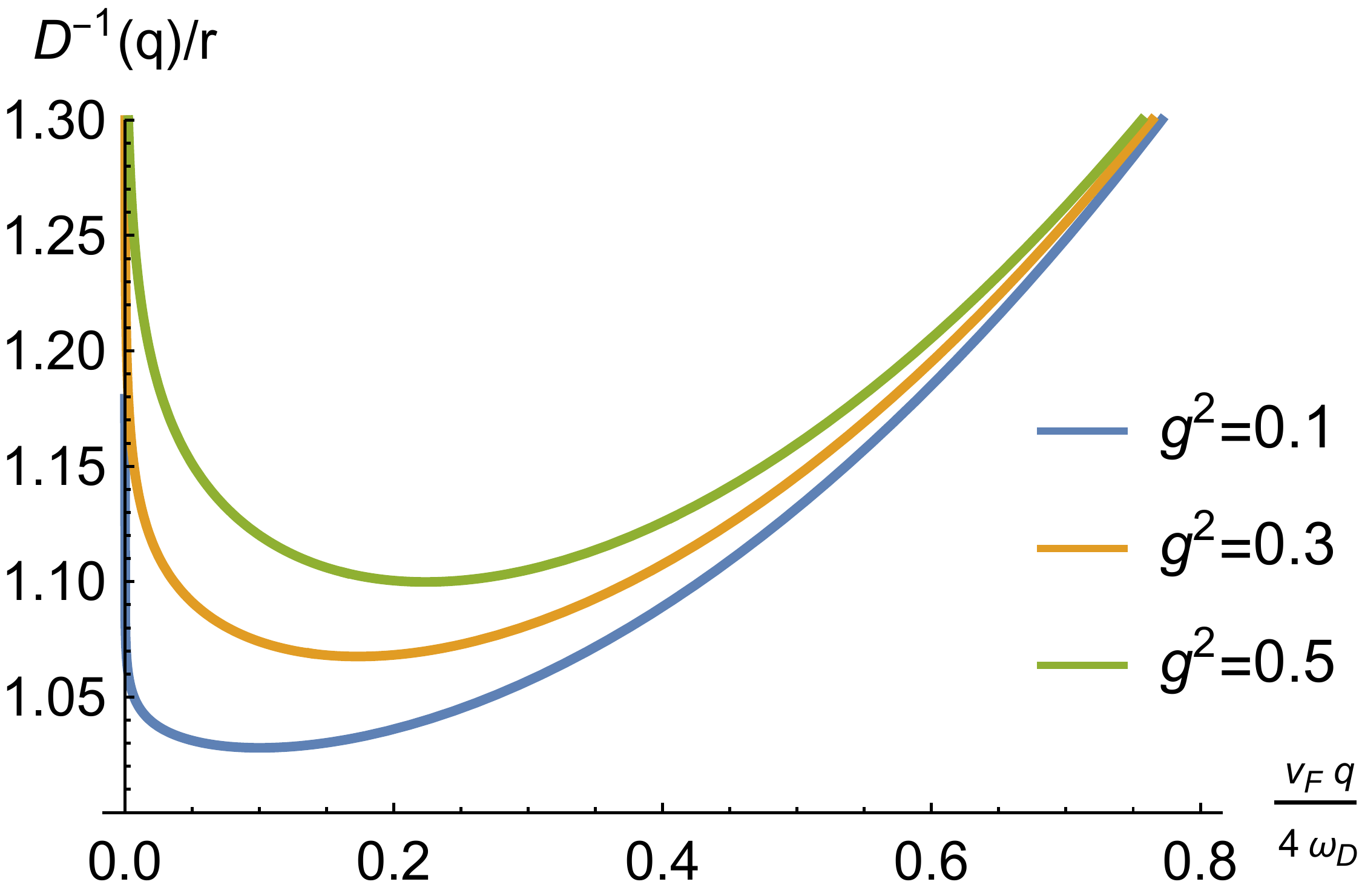}
	\caption{$D^{-1}(\bq)$ in the zero temperature limit obtained from Eq.\eqref{eq:D_inv1}. Here we set $c^2/r=0.5$, $\nu/r=0.1$, and the momentum is measured in units of $4\omega_D/v_F$.}\label{fig:flucPDW}
\end{figure}
We first show that when the interaction $V(r)$ is {\it monotonic}, e.g. $V(r) \sim e^{-r/\xi }$,  the PDW order is absent for any $g$.  The Fourier transform $V(\bm q)$ defines the bare inverse boson propagator, which is purely static, and takes an Ornstein-Zernike form: $D_0^{-1}(\bm q) = r + c^2 q^2$, with $r > 0$.  To see why the theory fails to host long range PDW order, consider the limit $q/2k_F \ll 1$, in which the saddle point solution for the exact static propagator $D$ at $T=0$ can be analytically obtained:
\begin{equation}
D^{-1}(\bm q) = r + c^2 q^2 + g^2 \nu \log{\left( \frac{4 \omega_D}{v_F q }\right) }, \label{eq:D_inv1}
\end{equation}
where the last term above is the contribution from the $q \ll k_F$ limit of the static pair susceptibility, $\omega_D$ is a cutoff, and $\nu$ is the density of states at the Fermi level.  Even at $T=0$, $D^{-1}(\bq)$ remains positive, indicating the absence of a phase transition.  Nevertheless, the minimum of $D^{-1}(\bm q)$ is at non-zero $|\bq|=\sqrt{\frac{g^2\nu}{2c^2}}$, indicating softened pair fluctuations at finite momentum.   Figure \ref{fig:flucPDW} shows $D^{-1}(\bm q)$ for various strengths $g^2$. With increasing $g^2$, the theory is driven further away from ordering, eventually having a correlation length short compared to the wavelength of the putative PDW - thus, a failed PDW.  We next show that long ranged PDW order occurs when the repulsive BCS couplings are non-monotonic in space.

{\it PDWs from non-monotonic BCS interactions.  } 
\begin{figure}
	\includegraphics[width=8.cm]{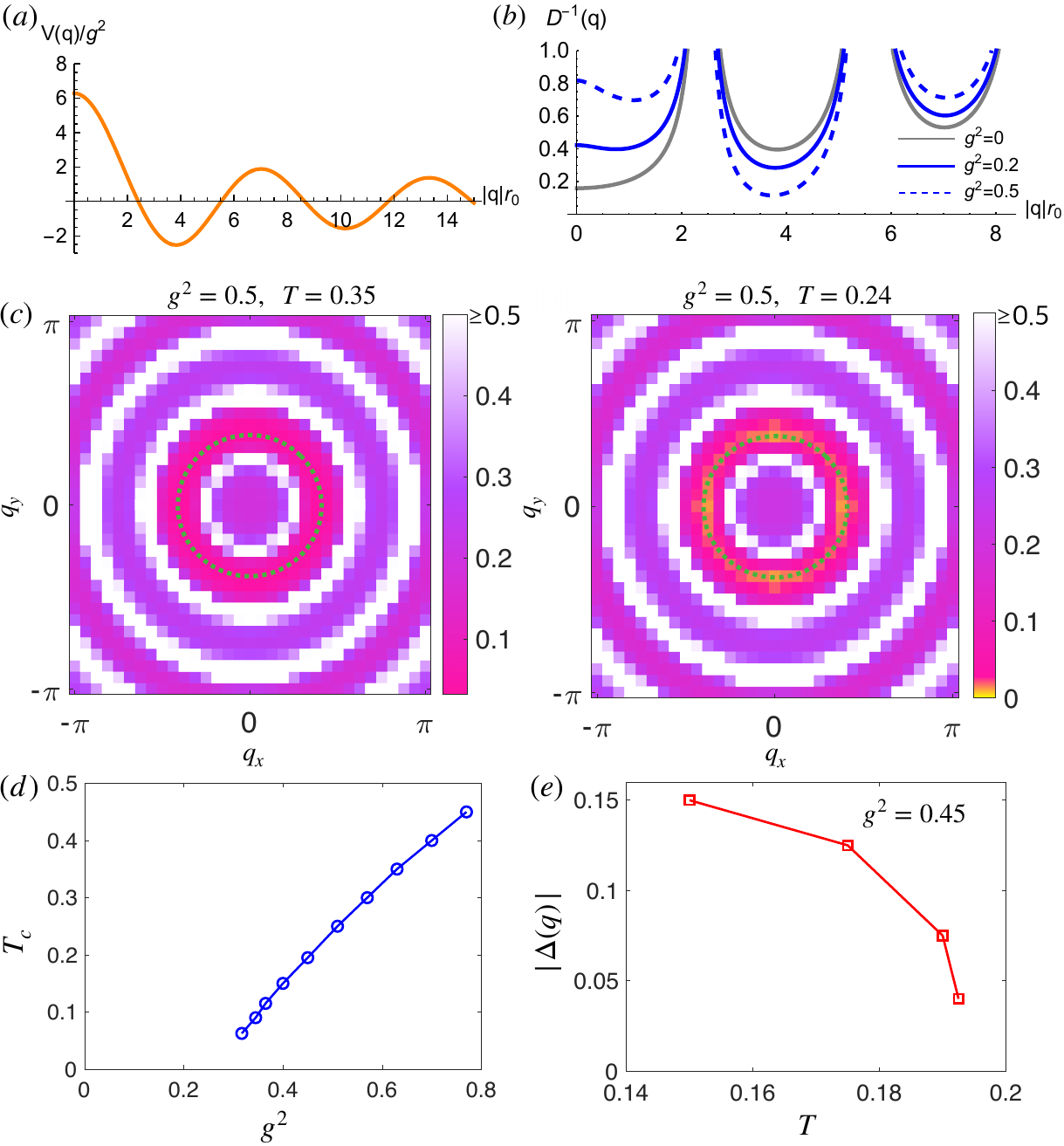}
	\caption{(a)$V(\bq)$ as a function of $|\bq|r_0$ with $r_0=1$ from Eq.\eqref{eq:toy}. (b) $D^{-1}(\bq)$ at $T=0.05$ as a function of $|\bq|r_0$ (also with $r_0=1$) obtained by approximating $\Pi(\bq)$ in Eq.\eqref{eq:Dq} by its one-loop calculation. (c) Density plot of $D^{-1}(\bq)$ as a function of $\bq$ obtained by numerically solving the full saddle point equations in \eqref{eq:saddle} with $r_0=3$. The two panels show the results for $T$ above $T_c$ and right at $T_c$, and the dashed circles mark the minimum of $D^{-1}(\bq)$. (d) $T_c$ as a function of $g^2$. At large $g^2$, our result indicate that $T_c$ scales linearly in $g^2$. (e) The magnitude of $\Delta(\bq)$ below $T_c$ for a given $g^2=0.45$. The energy scale here are measured in unit of the Fermi energy $E_F$.
	}\label{fig:PDWonset}
\end{figure}
As an illustrative example, consider the case where the BCS coupling is non-zero only at a distance $r_0$: 
\begin{equation}
V(\bm r) = g^2 \delta(r - r_0),  \ V(\bm q) = 2 \pi r_0 g^2 J_0(q r_0),\label{eq:toy}
\end{equation} 
where $J_0(x)$ is the zeroth Bessel function.  Although $V(\bm r)$ is repulsive, its Fourier transform $V(\bm q)$ is an {\it oscillatory} function with both repulsive and attractive components[Fig. \ref{fig:PDWonset}(a)].  
The exact boson propagator in this case is
\begin{equation}
D^{-1}(\bm q) = \frac{1}{2 \pi r_0 \vert J_0(q r_0) \vert} + g^2 {\rm sgn}\left[ V(\bm q) \right] \Pi(\bm q). \label{eq:Dq}
\end{equation}
To make sense of the above equation, we can approximate the boson self energy $\Pi(\bq)$ by the one-loop calculation $\Pi_0(\bq)$ obtained using $G_0$. The result is shown in Fig.\ref{fig:PDWonset}(b).
Clearly we see that when $V(\bm q)< 0$, the associated Fourier components of $D^{-1}(\bm q)$ get smaller (i.e. closer to an ordering transition) as $g^2$ increases whereas the repulsive components get larger.  Nonetheless, the phase transition will not occur unless $g^2$ exceeds a threshold value.   
In Fig.\ref{fig:PDWonset}(c) we present the numerical results of $D^{-1}(\bq)$ by solving the full saddle point equations \eqref{eq:saddle} on a $32\times32$ momentum mesh grid. The global minimum (dashed circle) of $D^{-1}(\bq)$ indeed vanishes when $T$ approaches $T_c$. 
Thus, there is a line of finite temperature phase transitions $T_c(g^2)$ as $g^2$ is varies, obtained by the condition $D^{-1}(\bm q) = 0$. For $T>T_c$, the minimum value of $D^{-1}(\bq)$ forms a ring as is expected from the toy model, but stays positive. Once $T$ approaches $T_c$, its minimum vanishes, indicating the PDW instability. Similarly, if we fix $T$ instead and increase $g^2$, we can also see $D^{-1}(\bq)$ vanishes at some finite $g^2$.   In Fig.\ref{fig:PDWonset}(d) we present $T_c$ and a function of $g^2$. At large $g^2$, our result clearly shows a linear relation between $T_c$ and $g^2$. The line of finite temperature transitions terminates at a $T=0$ phase transition at $g=g_c$.  

Below the ordering transition, we must solve the self-consistent equations allowing for a non-zero expectation value $\Delta(\bm q) = \langle \phi(\bm q) \rangle$.  Details of our calculation are provided in \cite{SM}. Fig. \ref{fig:PDWonset}(e) shows  $\Delta(\bq)$ as a function of $T$ below $T_c$.  Within the accuracy of the numerical solutions, the expectation value grows continuously indicating that the finite temperature transitions are second order and are well-decribed by mean-field theory.  From the solution of the non-linear equations, we can also determine the ordering wave-vector $\bm Q$ of the PDW by minimizing $D^{-1}(\bm q)$ with respect to momentum:
\begin{equation}
\bm Q: \frac{d}{d \bm q} D^{-1}(\bm q)\vert_{\bm Q} = 0
\end{equation}
In the neighborhood of $\bm Q$, $D^{-1}(\bq)$ takes the form $D^{-1}(\bm q) = \gamma \left(\textcolor{black}{|q| - Q }\right)^2$, where $\gamma = \frac{1}{2} \frac{d^2}{dq^2}D^{-1}(\bm q) \vert_{\bm q}$.

{\it Lattice models with PDW order.  }Emboldened by the simplified model above, we consider a more realistic example of electrons on a square lattice with nearest neighbor hopping $t=1$, onsite Hubbard $U$, and second neighbor pair-hopping $J$:  
\begin{equation}
	\begin{aligned}
		H=&-t\sum_{\braket{i,j},\sigma}c_{i \sigma}^\dagger c_{j \sigma}+U\sum_in_{i \spinup}n_{i \spindown}+J \sum_{\braket{i,j}}c_{i \spinup}^\dagger c_{i \spindown}^\dagger c_{j \spindown}c_{j \spinup},
	\end{aligned}\label{eq:lattice}
\end{equation}
where $i,j$ above label lattice sites.  The model above can similarly be N-enhanced and the resulting saddle point solutions can be solved {\it mutatis mutandis}.  
\begin{figure}
	\includegraphics[width=8cm]{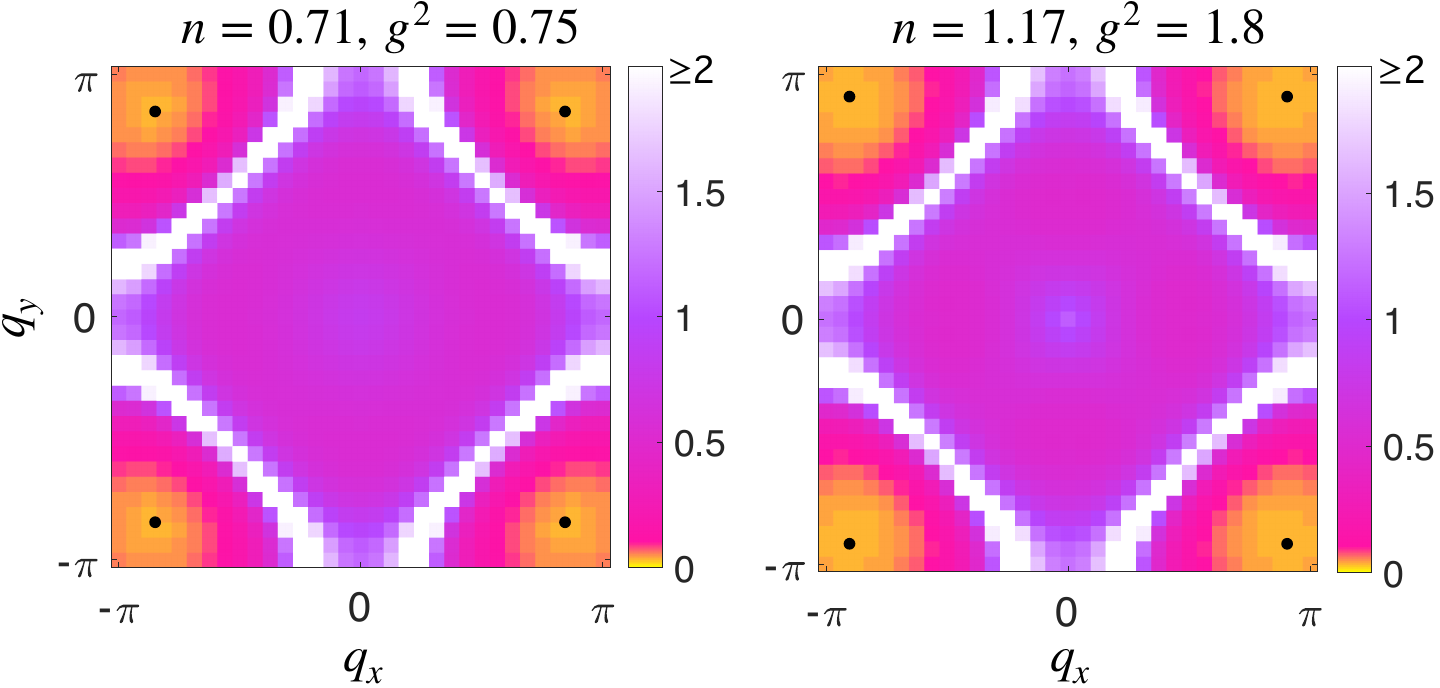}
	\caption{Numerical solution of $D^{-1}(\bq)$ from Eq.\eqref{eq:lattice} obtained at a fixed $T=0.05$ with $J=2U$ for different fillings. In the left panel, there are $n=0.71$ electrons per site, and $D^{-1}(\bq)$ touches zero at $g^2=0.75$. In the right panel, there are $n=1.17$ electrons per site, and $g^2=1.8$ is the critical coupling where $D^{-1}(\bq)$ touches zero. The black dots mark the positions of the ordering vector $\bQ$ near $(\pm\pi,\pm\pi)$, which leads to the PDW with checkerboard pattern in real space.}\label{fig:numerical}
\end{figure}
In this case, the Fourier transform of the BCS interaction $V(\bm q)$ is $V(\bm q) =U+2J(\cos q_x+\cos q_y)$ 
and $g^2 = U/t$. As long as $U<4J$, $V(\bm q)$ can be negative at some finite $\bq$. We  solve Eq.\eqref{eq:saddle} with the fermion dispersion replaced with $\xi_{\bk}=-2t(\cos k_x+\cos k_y) -\mu$. The results are shown in Fig.\ref{fig:numerical} In this case, we have four symmetry-related ordering vectors at $(\pm\pi,\pm\pi) + \mathcal O(U/J)$, that depend on the strength of interactions and the filling. In this sense, the pairing state from the large-$N$ theory is different from the $\eta$-pairing state found in numerical studies of one dimensional analogs of such models\cite{Yang1989,Yang1990,Hui1993,Japaridze_1997,Bhattacharyya_1995}.

{\it PDW quantum critical point.}
Both the lattice and continuum models above have finite temperature continuous PDW transitions that terminate at QCP.  We can study the fate of itinerant fermions around this $T=0$ transition by solving the self-consistent set of equations in \textcolor{black}{Eq.}~\eqref{eq:saddle}. 
A straightforward computation of the one-loop boson self-energy in the regime $q \ll k_F$ yields (see supplementary section \cite{SM})
$ \Pi(\bm q, i \Omega_m) = \nu(\ln\frac{4\omega_D}{v_F|\bq|}-\frac{|\Omega_m|}{v_F|\bq|})$.
 It then follows that in the limit $q \ll k_F$,  
 \begin{equation}
 D(q)^{-1} \approx \gamma(|\bq|-Q)^2+\frac{g^2\nu|\Omega_m|}{v_FQ},
 \end{equation}
  resulting in a boson dynamical exponent $z_b = 2$.  A fully self-consistent solution is obtained by computing the fermion self-energy using $D(\bm q)$ above.  Performing the integrals in the $z_b=2$ scaling limit (see supplemental sections \cite{SM}), we obtain $G^{-1}(\bm k, i \omega_n) = G_0^{-1}(\bm k)+  \textcolor{black}{ \Sigma(\omega_n)} $, where
  \begin{equation}
  \Sigma( \omega_n) = \textcolor{black}{i} {\rm sgn}(\omega_n) \omega_0^{1/2} \vert \omega_n \vert^{1/2}, \ 
  \omega_0 = \frac{g^2 Q}{\pi^2 v_F \gamma}.
  \end{equation}
  The expressions for $G,D$ are now fully self-consistent: upon feeding back the fermions to the boson, $\Pi$ is unchanged.  Thus, superconducting fluctuations are Landau overdamped and the fermions are dressed into a non-Fermi liquid.  If, following Hertz\cite{Hertz1976}, we were to integrate out the fermions, the bosonic sector would be at its upper-critical dimension defined by $d+z=4$, when $d=2$.  Thus, up to logarithmic corrections to scaling, the ordering transition has mean-field exponents, with $\nu=1/2$.  The line of finite temperature transitions emanates from the quantum critical point as $T_c(g^2) \sim \left( g^2 - g_c^2 \right)^{\nu z}$, with unit exponent.  
  Note that in our toy model Eq.\eqref{eq:toy}, the PDW ordering vector forms a ring, which renders the whole Fermi surface to be a `hot region'
  \footnote{Despite this peculiarity, fluctuation-induced first order transitions along the lines of Ref. \cite{Brazovski1975} are absent in the large $N$ limit since all dynamically generated non-linear terms in the boson effective potential are $1/N$ suppressed.  Nevertheless at finite N, we can argue against such first order transitions if we view the toy model in the continuum limit as applying to a dilute electron system on a lattice.  In this case, corrections to effective mass will lift the degeneracy and reduce the hot regions on the Fermi surface to hot spots.}.
   However, in the lattice model where there are only limited number of ordering vectors, there are only finite `hot spot' regions on the Fermi surface which has NFL behavior.

{\it Discussion. } We have shown that PDW order arises unambiguously when electrons have sufficiently large repulsive and non-monotonic BCS interactions.   Interactions in the particle-hole channel can certainly destabilize the theory presented here.  However, since  ordering tendencies in the particle-hole channel require finite interaction strength, we expect our theory to remain robust, at least to the addition of weak interactions in the particle-hole channel.  Other possibilities include Kohn-Luttinger superconductivity, which also arise from repulsive interactions.  However, such states are not present in the large $N$ limit considered here, and are moreover at exponentially low temperature scales; by contrast the PDW transitions occur at scales that exhibit power law dependence in the bare interactions of the system.


We speculate on the relevance of these results to real solids.  In microscopic descriptions of solids, pair-hopping interactions are typically small compared to density-density interactions\cite{Kivelson1987,Hirsch1989}.  This is not true, however, in low energy effective descriptions, obtained from integrating out short-distance modes.  In addition, it is somewhat unusual to expect a relatively suppressed BCS repulsion at short distances.  One possible manner to realize such suppression is to include strong coupling to Holstein phonons.  In such a strong coupling limit, the phonons induce instantaneous short-distance BCS interactions, which may help screen some of the bare short distance repulsion coming from say, a Hubbard interaction.    This may account for recent studies of Hubbard-Holstein ladders reporting PDW order\cite{Huang2022}.  A promising system for realizing the conditions outlined here for PDW formation are electrons on a Kagome lattice near the van Hove singularity.  In this regime, the electrons have a peculiar property that short distance Coulomb interactions are suppressed relative to nearest neighbor repulsion\cite{Kiesel2012}.  We shall investigate the possibility of PDW order in such models, and their relevance to the phenomena of Kagome metals such as CsV$_3$Sb$_5$, in future studies.

\begin{acknowledgements}
We thank D. Agterberg, A. Chubukov, R. Thomale, Z. Han, P. Hirschfeld, C. Murthy, S. Kivelson, and J. Sous for helpful discussions.  SR and PN were supported by the Department of Energy, Office of Basic Energy Sciences, Division
of Materials Sciences and Engineering, under Contract No. DE-AC02-76SF00515.  YMW acknowledges the Gordon and Betty Moore Foundation’s EPiQS
Initiative through GBMF8686 for support. 
AAP is supported by the Flatiron Institute. The Flatiron Institute is a division of the Simons Foundation.
 We thank the participants of the ICTP 2022 workshop on {\it Strongly correlated matter} for enjoyable discussions.  
 \end{acknowledgements}



\bibliography{PDW}

\afterpage{\null\newpage}

	\beginsupplement
	\begin{center}
  \textbf{\large  ONLINE SUPPORTING MATERIAL
  \\[.2cm] Pair density wave order from electron repulsion}
\\[0.2cm]
Yi-Ming Wu,$^{1}$ P. A. Nosov,$^{1}$ Aavishkar A. Patel,$^{2}$ and S. Raghu$^{1}$
\\[0.2cm]
{\small \it $^{1}$ Stanford Institute for Theoretical Physics, Stanford University, Stanford, California 94305, USA}

{\small \it $^{2}$ Center for Computational Quantum Physics, Flatiron Institute, New York NY 10010, USA}

  \vspace{0.4cm}
  \parbox{0.85\textwidth}{In this Supplemental Material we (i) comment on the decoupling of the repulsive BCS interaction, (ii) evaluate the one-loop pair susceptibility at finite external momenta and frequency (at any temperature), (iii) illustrate how we evaluate the magnitudes of the PDW gap function below instability temperature,  and (iv) compute the one-loop fermionic self-energy at the QCP.} 
\end{center}

\section{Hubbard-Stratonovich transformation for repulsive BCS interaction}
When the BCS interactions are repulsive, the effective action in Eq. \ref{eq:Lag} is not Hermitian.  We show here that there are no problems that arise from such an action.  
Consider the $0+0$ dimensional version of the theory (which is just an integral) to illustrate the point.  The generalization to the path integral is trivial. The identity we will use is 
\begin{equation}
e^{-U \hat O^{\dagger} \hat O} = \int \frac{d \phi d \phi^*}{2 \pi i} e^{- \phi \phi^* + i \sqrt{U} \phi^* \hat O + i \sqrt{U} \hat O^{\dagger} \phi} , \ \ \hat O = c_{\downarrow} c_{\uparrow}
\end{equation}
Treating $\phi, \phi^*$ as independent fields, their classical equations of motion are
\begin{equation}
\phi = i \sqrt{U} \hat O, \ \ \phi^* = i \sqrt{U} \hat O^{\dagger}
\end{equation}
at the classical saddle, $\phi^*$ is not the complex conjugate of $\phi$.  This naively seems problematic, since the integral over $\phi, \phi^*$ then appears divergent.  However, this is false.  To see why,
it is helpful to work in a manifestly real representation, defining 
\begin{equation}
\phi = x + i y, \ \ \phi^* = x - i y
\end{equation}
Then,
\begin{eqnarray}
 \int \frac{d \phi d \phi^*}{2 \pi i} e^{- \phi \phi^* + i \sqrt{U} \phi^* \hat O + i \sqrt{U} \hat O^{\dagger} \phi} = \int \frac{dx dy}{\pi} e^{-(x^2 + y^2) + i \sqrt{U} \left( \hat O + \hat O^{\dagger} \right) x  + i \sqrt{U} \left( \hat O^{\dagger} - \hat O \right) y}
 \end{eqnarray}
Now, the integrals over x,y can separately be done.  This is done by promoting each of them to a complex number, $x \rightarrow z_1, y  \rightarrow z_2$, and then deforming the contours appropriately to cross their respective saddle points, neither of which are on the real axis.  Thus, the decoupling is perfectly consistent.    This generalizes to the full functional integral. 

\section{Cooper pair susceptibility}
In this section, we revise the calculation of the one-loop pair susceptibility at finite external frequency and momenta 
\be \label{eq:Pi_c_def}
\Pi^{(c)}(q,i\Omega_m)=T\sum\limits_{n} \int\frac{d^2p}{(2\pi)^2} \frac{1}{(i\varepsilon_n +i\Omega_m-\xi_{p+q})(-i\varepsilon_n-\xi_{p})}
\ee
where $\varepsilon_n=2\pi T(n+1/2)$, $\Omega_m=2\pi T m$ are fermionic and bosonic Matsubara frequencies. We assumed some UV cut-off for Matsubara frequencies $\omega_{ D}$. First, we linearize the dispersion as $\xi_{p+q}\approx \xi_p +v_Fq\cos\theta$, where $v_F$ is the Fermi velocity, and $\theta$ is the angle between $q$ and $p$ (as usual, this approximation is sufficient for  $q\ll k_F$). After integrating over $\xi_p$, we perform the sum over Matsubara frequencies by means of the series representation of the digamma function. As a result, we arrive at the following expression for the pair susceptibility:
\be 
\Pi^{(c)}(q,i\Omega_m)=\nu \ln \left(\frac{2e^\gamma\omega_{D}}{\pi T}\right)-\nu \operatorname{Re}\left\langle\psi\left(\frac{\Omega_m+iv_Fq\cos\theta }{4\pi T} +\frac{1}{2}\right)-\psi\left(\frac{1}{2}\right) \right\rangle_\theta
\ee
where $\langle ...\rangle_\theta =(2\pi)^{-1}\int\limits_{0}^{2\pi}d\theta$ stands for the average over the Fermi surface, and $\gamma$ is the Euler's constant. Next, we can make use of the following integral representation of the digamma function
\be 
\psi\left(z+\frac{1}{2}\right) = \int\limits_{0}^{+\infty} dt \left(\frac{e^{-t}}{t}-\frac{e^{-zt}}{2\sinh t/2}\right)
\ee
After performing the remaining integration over $\theta$, we obtain
\be 
\Pi^{(c)}(q,i\Omega_m)=\nu \ln \left(\frac{2e^\gamma\omega_{D}}{\pi T}\right)-\frac{\nu}{2} \int \limits_{0}^{+\infty} \frac{dt}{\sinh t/2}\left(1-J_0\left(\frac{v_Fq t}{4\pi T}\right)e^{-\frac{|\Omega_m|t}{4\pi T}}\right)
\ee
After changing variables $t|\Omega_m|/(4\pi T)=z$, and defining dimensionless ratios $\alpha=2\pi T/|\Omega_m|$ and $\beta = v_Fq/|\Omega_m|$, we find
\be \label{eq:Pi_(c)_full}
\Pi^{(c)}(q,i\Omega_m)= \nu\left[ \ln \left(\frac{2e^\gamma\omega_{D}}{\pi T}\right)- \mathcal{K}(\alpha,\beta)\right]
\ee
where we introduced the following dimensionless function of $\alpha$ and $\beta$ 
\be 
\mathcal{K}(\alpha,\beta) = \alpha \int\limits_0^{+\infty}\frac{dz}{\sinh \alpha z}\left(1-J_0(\beta z)e^{-z}\right)
\ee
We emphasize that \eqref{eq:Pi_(c)_full} can be used at finite $T$ as long as $q\ll k_F$ (but for arbitrary ratio $\beta = v_F q/|\Omega_m|$).
Our next goal is to compute $\mathcal{K}(\alpha,\beta)$ in the limit $\alpha \rightarrow 0$ ($T=0$ limit), and for fixed $\beta=v_Fq/|\Omega_m|$. This can be done as follows. First, we observe that 
\be \label{eq:K_a0}
\mathcal{K}(\alpha,0) =\gamma +2\ln 2+\psi\left(\frac{1+\alpha}{2\alpha}\right)\approx \ln\frac{1}{\alpha}+\ln 2+\gamma+\mathcal{O}(\alpha)
\ee
for small $\alpha \ll 1$. Next, we differentiate $\mathcal{K}(\alpha,\beta)$ with respect to $\beta$ and find
\be 
\frac{\partial}{\partial\beta}\mathcal{K}(\alpha,\beta) = \alpha \int\limits_{0}^{+\infty}\frac{dz}{\sinh \alpha z}\; zJ_{1}(\beta z)e^{-z}
\ee
This expression has a well-defined limit $\alpha=0$:
\be 
\lim\limits_{a\rightarrow 0^+}\frac{\partial}{\partial\beta}\mathcal{K}(\alpha,\beta) =  \int\limits_{0}^{+\infty}dz J_{1}(\beta z)e^{-z}=\frac{1}{\beta}\left(1-\frac{1}{\sqrt{1+\beta^2}}\right)
\ee
Finally, we can integrate back over $\beta$ and use \eqref{eq:K_a0} as the initial condition 
\be 
\int\limits_{0}^\beta dy\; \frac{\partial}{\partial y}\mathcal{K}(\alpha,y) = \mathcal{K}(\alpha,\beta)-\mathcal{K}(\alpha,0)
\ee
After expanding both sides in small $\alpha$ and performing the remaining integral over $y$, we obtain
\be 
\mathcal{K}(\alpha,\beta) = \ln\frac{1}{\alpha} +\gamma +\ln\beta +\frac{1}{2}\ln\left( \frac{\sqrt{\beta^2+1}+1}{\sqrt{\beta^2+1}-1}\right) +\mathcal{O}(\alpha)
\ee
for small $\alpha\ll 1$. Finally, the full $T=0$ expression for $\Pi^{(c)}(q,i\Omega_m)$ takes the form
\be \label{eq:Pi_full}
\Pi^{(c)}(q,i\Omega_m)=\nu \ln\left(\frac{4\omega_{D}}{v_Fq}\right) +\frac{\nu}{2} \ln\left(\frac{\sqrt{|\Omega_m|^2+\left(v_Fq\right)^2}-|\Omega_m|}{\sqrt{|\Omega_m|^2+\left(v_Fq\right)^2}+|\Omega_m|}\right)\;,\quad T=0
\ee
with no assumptions on the ratio $|\Omega_m|/ (v_F q)$. If we now consider the limit  $|\Omega_m|\ll v_F q$, then we find
\be \label{eq:Landau_Damping}
\Pi^{(c)}(q,i\Omega_m)\approx \nu \ln\left(\frac{4\omega_{D}}{v_Fq}\right) -\frac{\nu |\Omega_m|}{ v_Fq}\;,\quad  |\Omega_m|\ll  v_Fq
\ee
We emphasize that exactly the same result can be easily obtained by taking the $T=0$ limit directly in the initial expression \eqref{eq:Pi_c_def}. Note that $\Pi^{(c)}(q,i\Omega_m)$ is related to the boson self-energy $\Pi(q,i\Omega_m)$ defined in the main text as $\Pi(q,i\Omega_m)= -g^2\operatorname{sgn}[V(\mathbf{q})]\Pi^{(c)}(q,i\Omega_m)$. 

\section{Green's functions below $T_c$} 
\label{sec:green_s_functions_below_}

The saddle point equations in \eqref{eq:saddle} can be used to extract the information on the PDW instability temperature $T_c$. To obtain the magnitude of the order parameter when $T<T_c$, we need to introduce the anomalous part of the fermion self energy, and search for convergent solutions. To this end, we add $g\sum_{k,\bq}(\psi_{\spinup}(k)\psi_{\spindown}(-k+q)\Delta(\bq)+\text{h.c.})$ into the parent action, such that the total action becomes
\begin{equation}
  \begin{aligned}
    \frac{S}{N}=&\int dr\left[-\eta^2g^2 G^2(r)D(-r)+ 2G(r)\Sigma(-r)+D(r)\Pi(-r)\right]\\
    &+\sum_{k,\sigma}\psi_\sigma^\dagger(k)[-i\omega_n+\xi_{\bk}-\Sigma(k)]\psi_\sigma(k)\\
    &+g\sum_{k,\bq}(\psi_{\spinup}(k)\psi_{\spindown}(-k+q)\Delta(\bq)+\text{h.c.})\\
    &+\sum_q\bar{\phi}(q)\left[|\lambda(\bq)|^{-1}-\Pi(q)\right]\phi(q)
  \end{aligned}\label{eq:largeNaction}
\end{equation}
Since this action is bilinear in both fermion and boson fields, we can integrate them out. For the fermion fields we work in Nambu space due to the presence of $\Delta(\bq)$, and the resulting action takes the form of $-\text{tr}\ln\mathcal{F}$, where $\mathcal{F}$ is a matrix given by
\begin{equation}
  \mathcal{F}_{\bk,\bk'}=\begin{pmatrix}
    [-i\omega_n+\xi_{\bk}-\Sigma(\bk,i\omega_n)]\delta(\bk-\bk') & \Delta(\bm{q})\delta(\bk'-\bk+\bm{q})\\
    \bar{\Delta}(\bm{q})\delta(\bk'-\bk-\bm{q}) & [-i\omega_n-\xi_{\bk}+\Sigma(-\bk,-i\omega_n)]\delta(\bk-\bk')  \\
  \end{pmatrix}\label{eq:kernelF}
\end{equation}
This matrix is diagonal in momentum space only if $\Delta(\bq)$ is has a finite value at $\bq=0$. This is not true for the PDW order, for which $\Delta(\bq)$ is finite at some finite $\bq=\bQ$. 

Then the routine procedure of variation leads to the same saddle point equations as Eq.\eqref{eq:saddle}, but with $G(k)$ modified into
\begin{equation}
	G(k)=\frac{1}{2}\text{tr}\left[\mathcal{F}^{-1}\frac{\delta \mathcal{F}_{\bk',\bk''}}{\delta\Sigma(k)}\right]\label{eq:F}
\end{equation}
At $T<T_c$ we use the solution at $T\geq T_c$ as input and solve Eq.\eqref{eq:F} by sweeping a set of values of $\Delta(\bq)$ with $\bq=\bQ$ and searching for the one which gives convergence after iteration.


\section{Fermionic self-energy}
In this section, we compute the one-loop fermionic self-energy induced by soft order-parameter fluctuations in a vicinity of a QCP at $g=g_c$. In the $T=0$ limit, we obtain
\begin{eqnarray}
\Sigma(\omega) &=&g^2 \int \frac{d^2 q d q_0}{(2 \pi)^3}  D(\bm q, q_0) G_0(\bm q - \bm k_F, q_0 - \omega) \nonumber \\
&=&  \frac{g^2}{(2 \pi)^3} \int d q_0 q d q d \theta \frac{1}{\gamma(q - Q)^2 + \frac{g^2 \nu \vert q_0 \vert}{v_F Q} } \frac{1}{i(q_0 -\omega) - v_F q \cos{\theta}} \nonumber \\
&=& -i\frac{g^2}{(2 \pi)^2} \int dq_0 q dq  \frac{1}{\gamma(q - Q)^2 + \frac{g^2 \nu \vert q_0 \vert}{v_F Q} }  \frac{{\rm sgn}(q_0- \omega)}{\sqrt{(q_0-\omega)^2 + (v_F q)^2}}\label{Sigma}
\end{eqnarray}
In the $z=2$ scaling limit, the typical frequency $q_0^2 \sim q^2\ll q^2$.  Therefore, the integral above can be approximated by
\begin{eqnarray}
\Sigma(\omega) &=&  \frac{-i g^2}{(2 \pi)^2v_F} \int dq_0 dq \frac{{\rm sgn}(q_0- \omega)}{\gamma(q - Q)^2 + \tilde g \vert q_0 \vert }, \ \ \ \tilde g \equiv \frac{g^2 \nu}{v_F Q}\nonumber \\
&=& \frac{-i g^2}{(2 \pi)^2 v_F} \int d q_0  \frac{{\rm sgn}(q_0 - \omega) }{\left(\gamma \tilde g \vert q_0 \vert \right)^{1/2}} \left[ \frac{\pi}{2} + \arctan{\left(\frac{\gamma^{1/2} Q}{\tilde g^{1/2} \vert q_0\vert^{1/2}} \right)} \right] \nonumber \\
&\simeq& \frac{-i g^2}{4 \pi v_F \left( \gamma \tilde g \right)^{1/2}} \int d q_0 \frac{{\rm sgn}(q_0 - \omega) }{ \vert q_0 \vert^{1/2}} ,   \ \ \ \ \ \left[ \gamma Q^2 \gg \tilde g \vert q_0 \vert \right]  \nonumber \\
&=& \frac{i g^2}{ \pi v_F \left( \gamma \tilde g \right)^{1/2}}{\rm sgn}(\omega) \vert \omega \vert^{1/2}
\end{eqnarray}
The self energy thus has the form
\begin{equation}
\Sigma(\omega) = {\rm sgn}(\omega) \omega_0^{1/2} \vert \omega \vert^{1/2} , \ \ \ \omega_0 = \frac{g^2 Q}{\pi^2 v_F \gamma \nu} \label{NFLSigma}
\end{equation}
This NFL behavior survives even beyond one-loop level. In fact, we can obtain $\Sigma(\omega)$ self consistently, by adding $\sgn(\omega)|\omega|^{1/2}$ in the fermion Green's function in Eq.\eqref{Sigma}, and following the same calculations as above, the result is still given by Eq.\eqref{NFLSigma}.

\end{document}